\newcommand{\LaSL}{LaTiO$_3$/LaCuO$_3$/LaTiO$_3$ }

\newcommand{\XSL}{XTiO$_3$/XCuO$_3$/XTiO$_3$ }

\documentclass[%
 amsmath,amssymb,
 aps,prb,twocolumn
]{revtex4-2}

\usepackage{amsmath}
\usepackage{graphicx}
\usepackage{dcolumn}
\usepackage{bm}

\begin{document}

\preprint{APS/123-QED}

\title{ Engineering superconducting properties of multiferroic copper oxide heterostructures}
\author{Wai Hei Terence Tse$^{1}$}
\author{Carla Lupo$^{1}$}
\author{Evan Sheridan$^{1}$}
\author{Evgeny Plekhanov$^2$}
\author{Cedric Weber$^{1}$}
\affiliation{%
 $^1$ King's College London, The Strand, London, WC2R 2LS, United Kingdom
}
\affiliation{
 $^2$ Quantinuum 13-15 Hills Road, CB2 1NL, Cambridge, United Kingdom
}

\date{\today}

\begin{abstract}
Oxide heterostructures have repeatedly been shown to display apical properties at the interfaces, some of which favorable to the formation of two-dimensional electron systems, as well as high transition temperature superconductivity. In this study, we propose a novel heterostructure to potentially achieve near room-temperature superconductivity, via the carrier injection in cuprate interfaces with ferro-electrics. Using a digital design approach guided by density-functional theory, the systems of \XSL are thoroughly examined, confirming the formation of a two-dimensional electron gas at the cuprous oxide interface. Via the manipulation of lattice parameters, the key ingredients for two-dimensional electron gas formation is explored. We apply cluster dynamical mean-field theory on the cuprous oxide plane and probe the superconducting properties of the system \XSL. As a result, we see a marked increase of superconducting ordering parameter near the fully occupied regime, a known marker for the superconducting transition temperature.
\end{abstract}

\maketitle

\section{Introduction}

Over the previous few decades, the superconducting properties of cuprous oxides have garnered intense focus and research. With the wide spectra of apical properties present, understanding the precise nature of the more unusual members of this oxide family promises novel new phenomena and pathways to future studies.

Strongly correlated systems, such as LaAlO$_3$/SrTiO$_3$ (LAO/STO) \cite{doi:10.1063/1.4901940}, a system made of two parallel planes of respective oxides. This structure was first proposed and described by A. Ohtomo and H. Y. Hwang in 2004, and unlike cuprate superconductors, the phenomena of interest exists within the interface between the two component layers. Within this region lies what is known as a high mobility quasi two-dimensional electron gas (2DEG) \cite{2DEG_LAO_STO}. The establishment of two-dimensional electron gas in this case is believed to be due to the polar-catastrophe mechanism \cite{2DEG}, and the presence of such is key to many of its atypical behaviors, one of the key features being planar interface superconductivity. Superconducting temperatures for these interfaces are usually observed at very low temperatures, and there is instability in the two-dimensional electron gas. However, there exists high transition temperature superconductors, or high Tc superconductors for short. Two-dimensional electron gas have also been observed between single layered, or thin film heterostructures \cite{Seok}.

Cuprates are layered materials constructed with layers of copper oxides separated with spacer layers, thus creating a multilayered structure. The spacer layers act as charge reservoirs, which in turn allows for superconductivity in the cuprous oxide plane. The structures in question can be described with the chemical formula $XS_{n+1}(CuO_2)_n$, where $n$ $CuO_2$ planes are 'sandwiched' and interlaced within $n+1$ layers of spacer planes S, X layers in depth. As copper has the strongest covalent bonding with oxygen, Cuprates are in a unique position among the numerous charge-insulating transition metal oxides (TMOs).

The cuprous oxide plane is insulated by spacer layers, and thus the resultant superconductivity phenomenon is confined within the cuprous oxide plane, exhibiting two-dimensional properties. Though well-known as a high-temperature superconductor, the exact mechanisms and properties that permit this high critical temperature are still under intense scrutiny today \cite{PhysRevX.8.021038}. One such superconductor of this variety is lanthanum copper oxide (La$_2$CuO$_4$), which is a well-known high-Tc superconductor, doped or otherwise \cite{La2CuO4}. Its behaviour is not well captured by the Hubbard model, however. 

Superlattices of magnetic transition metal oxides and LTO have also been investigated, in an attempt to control the magnetic behavior via the ferro-electric properties of lanthanum titanate \cite{LCoOLTO1,LCoOLTO2}. Here, the authors addressed the question of to what extent can cobalt magnetic electronic properties be controlled or manipulated by varying the local lattice structure surrounding the cation. 

Currently, there is a general clear consensus on the direction to obtain the room temperature superconductivity in cuprates, but the exact next steps to take is unclear. In our work, we propose a way to address this by building on existent theories. 

In recent studies \cite{Yayu,O'Mahony}, it has been realised that superconductivity cannot be fully captured with a one band Hubbard model, due to the copper-oxygen bonds that could be ionic or covalent. This directly affects the strength of correlations, which in turn affects Tc. It is realised that applying pressure metallizes the system by tuning the charge transfer energy \cite{Tremblay_2011}. However, to realize high Tc, we need to move the system towards the limit of ionic bonding. The difficulty arises from the necessity of negative pressure, which naively applying strain wouldn't accomplish.

Also, it has been shown that not only are there covalent and ionic regimes of Copper-Oxygen bonding, but its nature, and in addition Tc, are moderated by the charge transfer energy between Copper d orbitals and the $p_x$ and $p_y$ orbitals of Oxygen. Furthermore, it has been suggested that higher Tc can occur at the limit of weak coupling between the aforementioned orbitals \cite{Kowalski}. Here, we offer a realisation of this weak coupling (small hopping parameter $t_{dp}$) via epitaxial engineering.

By interfacing cuprates with ferro-electrics, it is possible to tune the properties of the two-dimensional electron gas interface system, and explore doping domains which are previously not explored. To tune the system, we turn to applying epitaxial strain by spacing out the atoms of the substrate the system is grown on. By changing the substrate, we can tune the strength of correlations in the direction we desire \cite{Carla_Evan}. Suggestions to realize this experimentally have also been made. With this method, we access a regime not accessible by chemical doping, where lanthanum copper oxide is electron-doped from charge transfer due to ferro-electricity, creating a two-dimensional electron gas. By combining two-dimensional electron gas generation on thin film copper oxide via ferro-electricity, and tuning correlation strength via epitaxial strain, we aim to create a new platform on which new systems with improved high temp superconductivity can be derived from known high Tc conductors. In particular, we are interested in how superconductivity charges with control parameters and providing a framework to test this hypothesis to understand superconductivity. In addition, we note that LaTiO$_3$ is a well known ferroelectric, and Copper doping of LaTiO$_3$ via Lanthanum substitution \cite{Chen} or Titanium substitution \cite{Sotelo} has been achieved experimentally within the field of catalysis. Similarly, oxide interfaces such as LTO/STO have also been studied \cite{Ohstuka}, and the concept of charge transfer from LTO to an interface 
and the creation of a two-dimensional electron gas is well known. The realisation of a two-dimensional electron gas can be obtained even up to room temperature, as obtained in tri-colour ferro-electric stacks recently \cite{Cao}.

Hence, we introduce a concept that addresses the recent findings on how superconductivity can be optimized, combined with the concept of tuning strength of correlation with epitaxial strain. Building on previous work which have given guidance to the design \cite{Cedric_1,Cedric_2}, we introduce a novel system \XSL; an interfacing of ferro-electrics to copper layers, which induces a superconducting phase within the ionic bonding limit. 2DEGs move freely in two dimensions while being quantized in the third, and the 2DEG present in interfaces displays superconductive properties and other novel phenomenon. By using single layer (thin film) cuprous oxide as a key component in a layered system à la LAO/STO, one could reasonably expect to observe novel phenomena along key layer interfaces, such as high Tc. 

There have been both experimental and theoretical works that highlight the controllable nature of in-plane charge transfer energy of certain systems via structural manipulation; and in addition, this energy is able to be directly related to Tc in cuprates \cite{Yayu_2}. Trembley et al. have found that the nature of bonding plays an additional role in controlling super-exchange interactions. Charge-transfer super-exchange is the electron-pairing mechanism of many superconductors, and is also controlled by the charge transfer energy \cite{O'Mahony}. The direct relation between Tc, charge transfer energy and super-exchange has also been observed experimentally. In recent state-of-the-art experiments. We build upon this very recent body of work to use identified guidelines for controlling Tc via epitaxial engineering. Interfacing oxides has long been known to provide a pathway for tuning properties. In the same vein, we propose a novel mechanism that combines ferro-electricity and induced metallicity at the oxide interfaces to induce charge carriers and control the distance of copper and apical oxygen.

In this paper, we explore the \XSL systems in detail with a focus on \LaSL
by using a conjunction of density-functional theory (DFT), and cluster dynamical mean-field theory (DMFT). We thus provide theoretical evidence of formation of a two-dimensional electron gas induced by anti-ferroelectric order in interfaced LTO/LCO stack systems. Through an examination of ground state energies of the suppressed structure, a tendency of the non-polar structure towards anti-ferromagnetism is found, while the polar structure is found to be non-magnetic. A shifting of lattice positions also reveals that lanthanum-oxygen polarisation plays a significant factor in 2DEG formation and quenching. 

To extract the cuprous oxide Anderson impurity model, the system is downfolded onto a three-band cuprous oxide band model via maximally-localised Wannier functions \cite{Wannier90} and downfolding. Finally, we carry out a cluster DMFT study of the superconducting order parameter and investigate the pairing symmetry of the superconducting phase at the interface. We obtain two regimes - a moderate superconducting order parameter in the vicinity of an anti-ferromagnetic phase with Cu-d9+$\delta$, which corresponds to low electron doping, and a largely enhanced superconductivity in the highly electronically coped limit, which corresponds to the stable structure. The \LaSL stack, when under induced charge transfer and anti-ferroelectric order, with large compression of the Copper - apical oxygen bonds, provides a pathway to achieve the novel superconducting phase at high electronic doping. We note that electron doped LCO can be achieved by chemical doping, such as in the work of Tsukada et al. \cite{Tsukada}. 

\section{Methods}

\subsection{Density Functional Theory (DFT)}

Relaxation of the \XSL system were carried out using the ab-initio DFT solver VASP \cite{VASP1,VASP2,VASP3} with PAW pseudopotentials \cite{VASP4,VASP5} and using the linear gradient method (LDA). In LDA, the exchange-correlation depends only on the local electron density.

In addition, the DFT+U method is used. This method introduces an additional Hubbard-like term that accounts for the strength of the on-site interactions, described by parameters U (for Coulomb interactions) and J (for exchange interactions). Here, we use the LDA+U variant \cite{Himmetoglu}. 

In the Dudarev derivation, $U$ incorporates the exchange parameter in the form $U_{eff}=U-J$, known as Hubbard U \cite{Dudarev}. 

In order to solve the 3-band Hamiltonian,

\begin{equation}
H=\sum_{i{\alpha}j\beta\sigma}t^{\alpha\beta}_{ij}c^{\dagger}_{i\alpha\sigma}c_{j\beta\sigma}+\sum_{i\alpha\sigma}\epsilon_{\alpha}n_{i\alpha\sigma}+U_{dd}\sum_{i\sigma}n_{id\uparrow}n_{id\downarrow}
\end{equation}

We employ a 2$\times$2 cuprate cluster Anderson impurity model. For 4 Cu impurity sites, the lattice Green’s function reads as,

\begin{equation}
\boldsymbol{G}_{\boldsymbol{k}}(i\omega_{n})=(i\omega_{n}+\mu-\boldsymbol{H}_{\boldsymbol{k}}-\boldsymbol{\Sigma}(i\omega_n))^{-1}
\end{equation}

where $\boldsymbol{H}_{\boldsymbol{k}}$ represents the Fourier transform of the uncorrelated part of the 3-band Hamiltonian, and $\boldsymbol{\Sigma}$ the self energy matrix of the cluster.

To solve for $\boldsymbol{\Sigma}$, we find the solutions for the 2$\times$2 AIM. The DMFT self-consistency condition is imposed:

\begin{equation}
i\omega-E_{imp}-\boldsymbol{\Sigma}(i\omega)-\boldsymbol{\Delta}(i\omega)=\hat{P}\left(\sum_{\boldsymbol{k}}G_{\boldsymbol{k}}(i\omega)\right)
\end{equation}

Here, the AIM can be defined as:

\begin{multline}
H_{imp}=\sum_{mn\sigma}\left(\epsilon^n_{mn\sigma}a^{\dagger}_{m\sigma}a_{n\sigma}+\epsilon^a_{mn\sigma}a^{\dagger}_{m\sigma}a^{\dagger}_{n-\sigma}+h.c.\right)\\ +
\sum_{mi\sigma}V_{mi\sigma}\left(a^{\dagger}_{m\sigma}c_{i\sigma}+h.c.\right)+\mu\sum_{i\sigma}c^{\dagger}_{i\sigma}c_{i\sigma}+\sum_{i\sigma}U\hat{n_{i\downarrow}}\hat{n}_{i\uparrow}
\end{multline}

The fermionic creation operators $a^{\dagger}_{m\sigma}$ and $c^{\dagger}_{i\sigma}$  creates a particle in the bath and in the cluster of impurities respectively, whereas the annihilation operators $a_{m\sigma}$ and $c_{i\sigma}$ destroys a particle accordingly.

This is summated over the bath sites indicated by indices m and n, while the index i indicates the impurity sites. Long-range hopping matrix elements connect the sites of the bath through via the particle-hole channel $n$ and particle-particle channel $a$. Non-correlated sites of the bath connect to the correlated impurities via matrix elements $V_{mi}$. Lastly, the onsite repulsion at the impurity sites is $U$, and the chemical potential is $\mu$. 

The resulting cDMFT AIM is solved as described by the methods of ref. \cite{Weber_2012}.

For the determination of the phase diagram we measured the S-wave and D-wave superconducting order parameter, being $\Delta_S=\frac{1}{2}(c_{12}+c_{14})$ and $\Delta_D=\frac{1}{2}(c_{12}-c_{14})$ respectively, derived from physical observables readily available from the 2$\times$2 impurity cluster. In indices in $c_{ab}$ represents the sites a and b, with the impurity sites being labelled anti-clockwise from 1 to 4.

The plotting tool sumo was used for certain figures \cite{sumo}.

\section{Results}

\begin{figure*}
\centering
\includegraphics[width=\textwidth,height=1.0\textwidth,keepaspectratio]{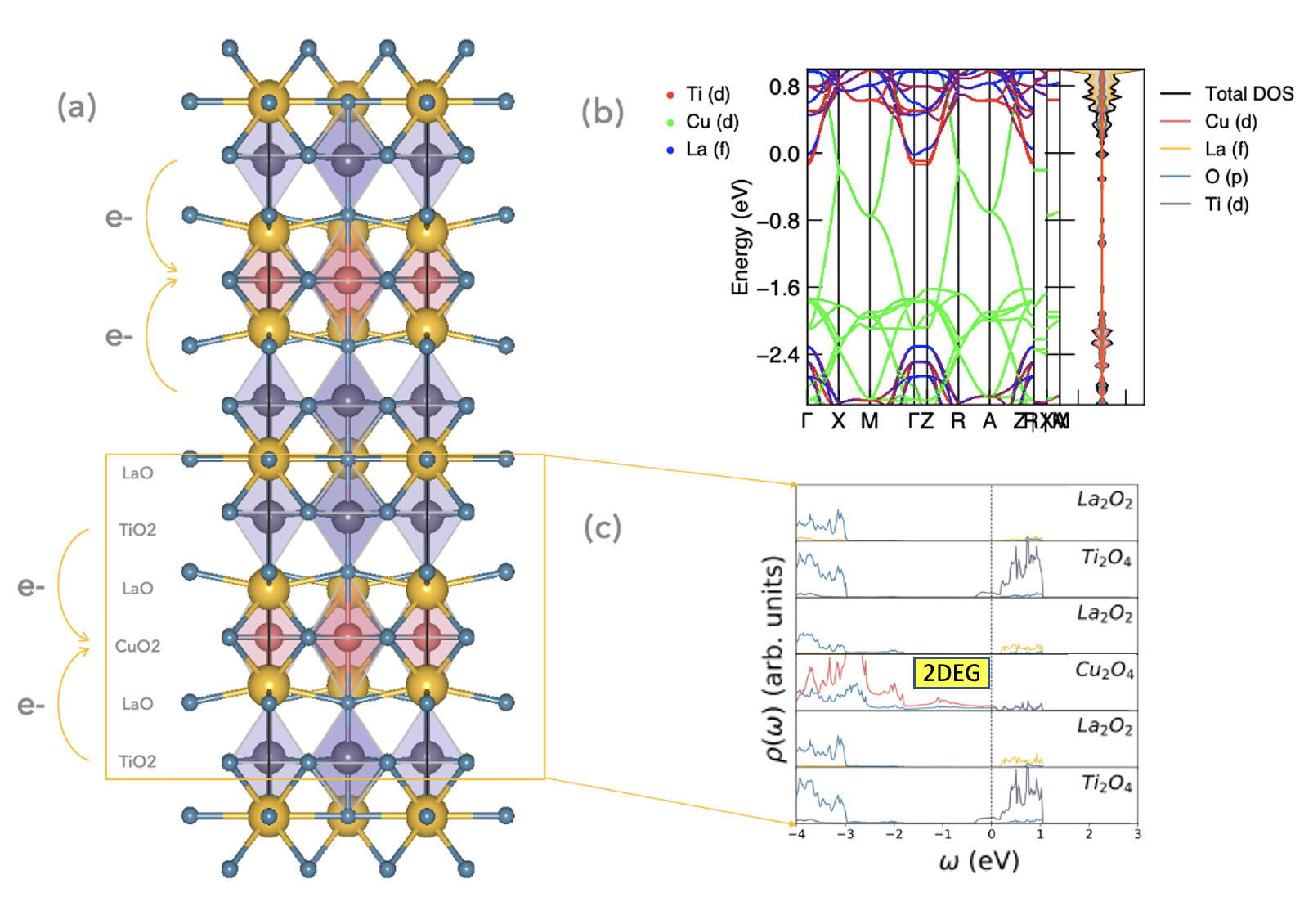}
\caption{Atomically resolved layer by layer density of states (c) for the Lanthanum superlattice with a +U correction on the copper and titanium states, with lattice structure (a) for the parent Lanthanum superlattice as reference. The color of elements in the structure corresponds to the elements shown in the DOS, and arrows on the left represent the direction of electron transfer. For this and all subsequent depictions of DOS, the Fermi level is shown with a dashed line, and equivalent to 0 eV. (b) Band structure of titanium, oxygen, copper and lanthanum (left), with corresponding total dos as reference (right).}
\label{fig:La_layer}
\end{figure*}

\subsection{The \XSL System}

From fig.\ref{fig:La_layer}, the structure in question can be disseminated clearly. As the figure suggests, \LaSL is a perovskite-like superlattice. This structure is equivalent to a $1\times1\times3$ supercell of LaTiO$_3$, with Cu ions in place of Ti in the second and third b-site layers. As a result, there are six distinct octahedral layers within the superlattice unit cell, differing only by octahedral central atom and/or octahedral orientation. In this construction, CuO$_2$ acts as the drain layer, TiO$_2$ the pump layer, and LaO the buffer layer. Consequently, TiO$_2$ layers will donate electrons to CuO$_2$, and The resultant charge transfer causes doping via ferroelectricity on the CuO$_2$ layer, creating the novel phenomena of interest. 

As we are looking for a system that can be scaled up to bulk, rather than just a thin film, we sandwich this system with layers of LTO/LCO/LTO. Otherwise, there is an issue where ferro-electricity causes polar catastrophe; as layers build up the system tends to a metal as ferro-electric is a potential that increases as layers increase. As a result, chemical potential shifts, and we end up with a trivial metal. To validate that the resultant system will be antiferro-electric and thus net polarization cancels out, DFT is used.

The total density of states, plotted as a function of energy are calculated using density functional theory with the plane-wave basis code VASP \cite{VASP1,VASP2,VASP3} using the projector augmented wave (PAW) PBE potentials \cite{VASP4} without spin-orbit coupling (SOC), with the energy cutoff of 500 eV, a 5x5x2 $\Gamma$-centered k grid, and a Gaussian smearing of 0.02 eV.

With DFT, the structure of the superlative is relaxed. Here, both lanthanum and titanium polarise to enable charge transfer, with a La-O polarisation of 0.45478\AA between TiO$_2$ and CuO$_2$, and non-polarised between CuO$_2$ layers. The atomically resolved layer by layer density of states (DOS) is found to be as fig.\ref{fig:La_layer}. The DOS shown are of the Ti-d, Cu-d, La-f and O-p variety, with respective colors corresponding to fig.\ref{fig:La_layer}(a). From the DOS, a broad band can clearly be seen with a bandwidth of 2.0 eV, which are constrained to their respective Cu-oxide $x$-$y$ planes. This feature is indicative of 2DEG in the drain layers of the superlattice and is mediated by the ability of the La3+ ion to induce a partial charge transfer of titanium from its 3+ oxidation state to 4+. This serves as a confirmation of the formation of a 2DEG in the cuprate interface, indicating the existence of the novel features discussed.

In addition, mixed charge carriers can be seen at the Fermi level (Ti-d and Cu-d). There are two types of band alignment, where O-p states align in Ti layers or O-p states align in La/Cu layers. Also, strong Cu-O hybridization can be seen at and below the Fermi level, and the copper bands are split into lower and upper Hubbard bands. As a whole, the structure is non-magnetic, and is metallic in titanium and copper.

\subsection{Magnetic to Paramagnetic Transition Driven by Charge Transfer}

Under anti-ferromagnetic and non-magnetic preparations, DFT is performed without prior relaxation, which allows the most favoured state of the non-relaxed (non-polar) structure to be found.

The free energies of each magnetic configuration are presented here in the table \ref{Tab:MAGMOM}. From the results, it is clear that DFT+U predicts that the anti-ferromagnetic state is the lowest in energy, whereas for DFT, the non-magnetic state has the lowest energy by a marginally small amount. This implies that under non-polar conditions, the system will tend towards an anti-ferromagnetic state. However, it is important to note when the system is allowed to relax and thus to polarise, this behaviour is not manifested, and the system reverts to being non-magnetic.

\begin{table}[h]
          \begin{center}
                \footnotesize
\begin{tabular}{ |p{3cm}|p{1.5cm}|p{1.5cm}|}
     \hline

 \multicolumn{3}{|c|}{Non-polar stack energy (in eV per cell)} \\
 \hline
 \hline
 Method & NM  & AFM-G \\
 \hline
      \hline
 DFT+U   &-221.484 & \textbf{\underline{-221.488}} \\
 \hline
 DFT     &\textbf{\underline{-238.363}}&-238.362\\
 \hline
                \end{tabular}
                  \end{center}
        \caption{Free energies of 30-atom unit-cell calculations in the non-polar superlattice conformation, subject to different magnetic configurations, calculated using DFT and DFT+U. The magnetic configuration with the minimum energy for each method has been underlined. DFT+U stabilises an anti-ferromagnetic phase in the metastable non-polar conformation (in the polar configuration, upon charge transfer, the system becomes paramagnetic).
        }
\label{Tab:MAGMOM}
\end{table}

\subsection{Quenching of 2DEG via Polarization}

As charge transfer within the system depends heavily on La-O polarization, elimination of such polarization in the system could stop charge transfer from occurring. Here, a comparison is made between the non-polar structure and polar structure, where the structure is shifted from its initial non-polar state, with 0 polarisation, to its final polar state of 0.52088\AA, identical to the relaxed state structure. This is reflected by the shortening and eventual elimination of the Cu broadband at the Fermi energy.

In addition to 2DEG quenching, a free energy difference of 4.656 eV between the non-polar and relaxed polar structure is found. 

The significance of this is twofold. Firstly, it shows explicitly that the 2DEG phenomena observed is dependent on the charge transfer between layers. Moreover, as the polarization of the layers can be manipulated externally, such as via application of potential difference, this implies that the superconductivity of the system can be influenced externally, unlike chemically doped cuprate superconductors, this system is ferro-electrically and structurally doped, and thus controllable via external parameters. This fact allows for the system to change superconducting behavior in a controllable fashion.

\begin{figure}[h]
\centering
\includegraphics[width=0.4\textwidth]{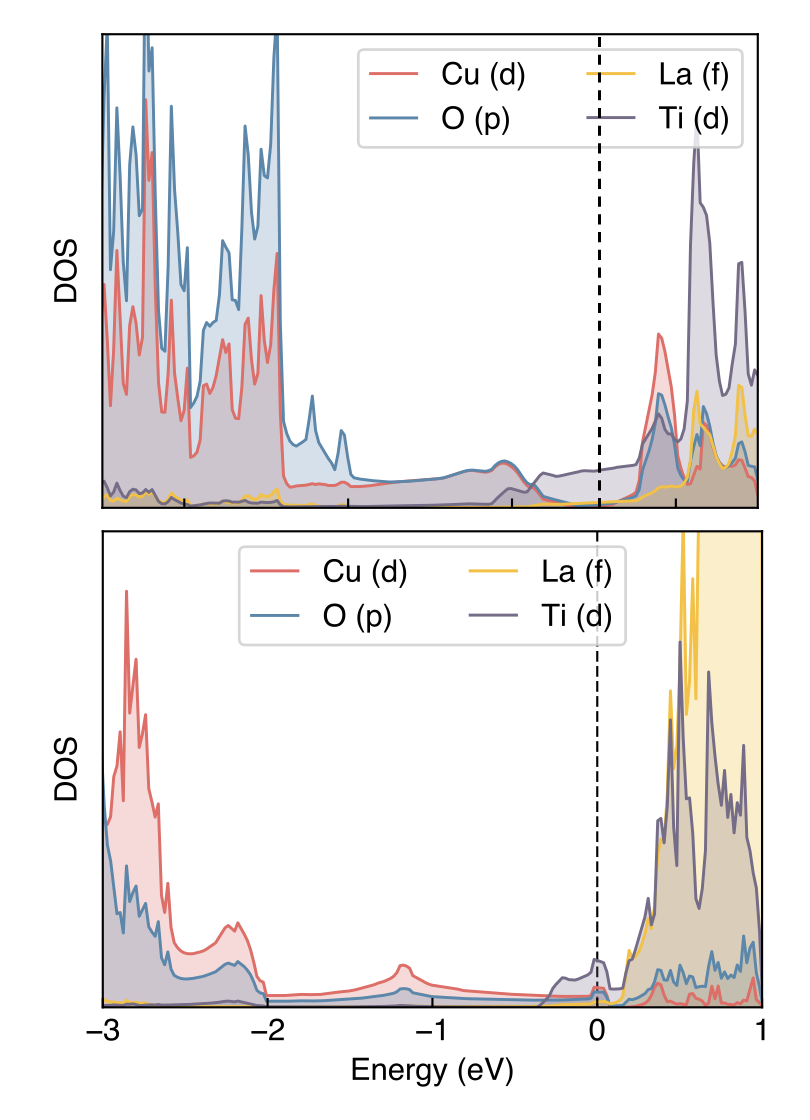}
\caption{\label{fig:dft}The atomically resolved DOS of AFM non-polar LCO/LTO (top) and non-magnetic polar (relaxed) LCO/LTO (bottom). As predicted, the width of 2DEG broadband at the Fermi level decreases with polarisation and eventually ceases as polarisation is removed, creating a gap.}
\end{figure}

\section{Electric Correlation and Charge Transfer}

\subsection{Obtaining an Effective Theory with Downfolding}

With the confirmation of superconductivity within the interface, we turn to probe its nature in the \XSL 2DEG. As the 2DEG is localised within the Cu layer interfaces, a 2x2 2-impurity model can be constructed, with Copper on each site, connected to two impurities corresponding to the adjacent oxygen. To accomplish this, it is optimal to reduce the current 30-atom system to 15-atom, such that there are exactly 1 Cu atom and 2 O atom in the CuO$_3$ layer per unit cell.

To construct the impurity c-DMFT model, the model Hamiltonian of the CuO orbitals are needed. Using Wannier functions, the system is then projected onto 13 bands, the lowest amount of projected bands realised that also retains the band features of the system around the Fermi level. 

By employing Wannier90 \cite{Wannier90}, maximally-localized Wannier functions (MLWFs) are produced and utilised to extract the hopping parameters, which are then downfolded onto the 3-band cuprate model. As downfolding sacrifices atomic character for the accuracy of bands at the Fermi level. The bands within the frozen window are compared with bands generated via VASP, showing good correspondence. For the 13 band model, $d_(x^2-y^2)$ of copper and all $p$ of apical and in-layer oxygen orbitals are projected upon; as for the 19 band model, $d_(x^2-y^2)$ of copper and the $p_x$ and $p_y$ orbitals of all oxygen are projected on. The resultant WFs are treated with 1000 iterations of disentanglement and 5000 iterations of spread minimisation. The inner window and frozen window are chosen as 16eV and 2eV respectively, centred on the Fermi level.

\begin{figure}
\centering
\includegraphics[width=0.4\textwidth,height=0.4\textwidth,keepaspectratio]{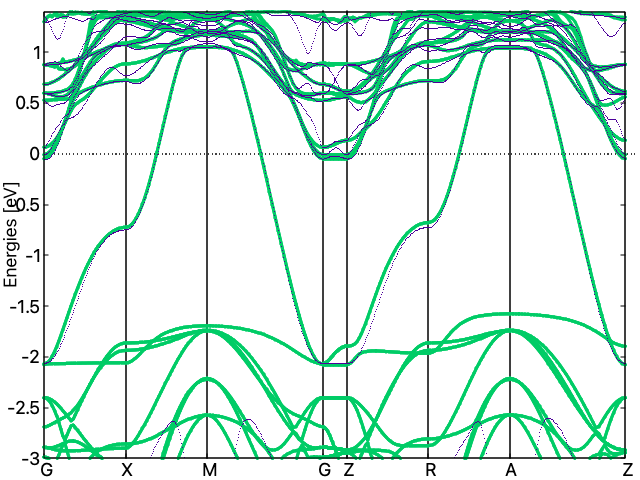}
\caption{Exaggerated comparison of resultant projected band structure from Wannier90 (Purple) and VASP (Green), showing good correspondence at the Fermi level.}
\label{fig:wannier_comp}
\end{figure}

Thus, we obtain the corresponding hopping parameters of the aforementioned 13-band projection, from which we extract the model Hamiltonian parameters via downfolding. 

Additionally, the same procedure is repeated for the 19-band projection of the Lanthanum structure, and a 13-band projection of the Yttrium substituted structure for comparison. From the projections taken, it is found that the 2p orbitals of oxygen apical to the Cu atom are essential for the precise fitting of Wannier bands, in addition of the orbitals within the CuO layer. We then proceed to downfold onto a 3-band Hamiltonian, which describes the 3dx2-y2 orbital of Copper and 2p orbitals of Oxygen within the copper oxide layer. Up to the second order hopping, we obtain all Cu-O and O-O hoppings in a 5$\times$5 lateral supercell, centred on the origin unit cell. It is important to note that the magnitude of hoppings are small relative to other oxide heterostructures, which as we elaborated prior is a good indicator of higher Tc.

\subsection{Many-body Theory of Superconductivity}

Using an ED impurity solver, the ordering of parameters of the 3-band model can be found.

The Hamiltonian of the three-band cuprate model can be expressed as:

\begin{equation}
H=\sum_{i{\alpha}j\beta\sigma}t^{\alpha\beta}_{ij}c^{\dagger}_{i\alpha\sigma}c_{j\beta\sigma}+\sum_{i\alpha\sigma}\epsilon_{\alpha}n_{i\alpha\sigma}+U_{dd}\sum_{i\sigma}n_{id\uparrow}n_{id\downarrow}
\end{equation}

Where $t^{\alpha\beta}_{ij}$ represent the hopping parameters from orbital $\alpha$ in unit cell i to orbital $\beta$ in unit cell j.
\begin{figure*}
\centering
\includegraphics[width=\textwidth,height=1.0\textwidth,keepaspectratio]{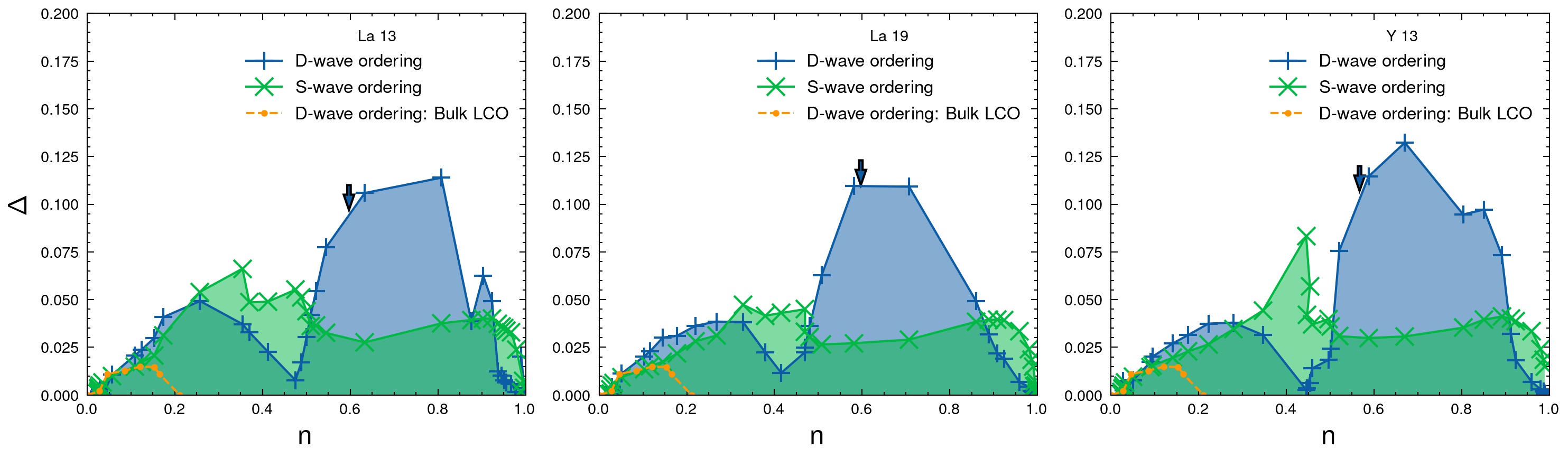}
\caption{d-wave (crosses) and s-wave (stars) components of $\Delta$, the superconducting order parameter.  $n$ is the carrier density, $n=0$ for the undoped compound corresponding to a Cu-d9 configuration, and $n=1.0$ is the full Cu-d10 atomic shell. The vertical arrow indicates the electron doping corresponding to the physical polar heterostructure.}
\label{fig:impurity}
\end{figure*}

Our results, presented in Fig. \ref{fig:impurity}, show the ordering of parameters for the downfolded Lanthanum 13-band, Lanthanum 19-band, and Yttrium 13-band model respectively. The characteristic dome of d-wave superconductivity can be seen at the ranges of 0.00 to 0.45, with a secondary peak at ~0.60. Between 0.35 to 0.5, S-wave superconductivity dominates. The calculations were performed at an inverse temperature equivalent to T $\approx$ 35K, and at U = 5eV.

We find phase transitions between S-wave and D-wave superconductivity as doping is increased, with an marked increase of order parameter in the D-wave superconductivity regime near the doped limit. The results suggest that two d-wave superconductive regimes, separated by an s-wave superconductive dominated state. From the order of parameters, we can see the d-wave superconductive regimes near the fully doped limit has a dome with peak value two to three times higher than the other.

\section{Conclusion}

We now turn to discuss the significance of the findings. Through DFT+DMFT, the properties of the novel \XSL system were explored. Using both DFT DFT+U, 2DEG phenomena was observed at the cuprous oxide interface due to inter-layer charge transfer, which in turn is meditated by lanthanum polarisation. In the stressed state, we found that it is indeed magnetic and insulating, with no charge transfer. In conclusion, we found a system in its ground state, with unnatural LCO which is doped and metallic. Through AIM c-DMFT, we find two distinct doping regimes of D-wave superconductivity, separated by a region of S-wave superconductivity.

With our current method, it is difficult to determine the exact Tc; however, by comparing the maxima of the two domes present in the doping domain, we see a remarkable increase in maximum order of parameters in the regime near the doped limit. For similar materials, Tc is found to scale with order of parameter, so a large increase of order of parameter would all but guarantee a significant increase of Tc. Our proposed LCO/LTO heterostructure yields a multi-fold increase of the maximum pairing order parameter, as compared to bulk LCO. Although the maximum pairing parameter does not correlate unequivocally with Tc near the integer filling regime, at large doping it provides an excellent marker for proportional changes in superconducting temperatures \cite{Weber_2012}.

Therefore, we have engineered theoretically an electron doped LCO high Tc. Although done chemically in the past, we show that using charge transfer, induced by stacking with polar interfaces, it is possible to generate 2DEG in the LCO copper oxide plane with Copper near the d-10 boundary. The 2DEG occurs in the limit of ionic bonding, as the Copper to Oxygen hybridisation is weak in this regime. This in turn leads to high Tc, and opens new avenues for designing superconductors by interfacing ferro electric materials with transition metal oxides in the future.

The structure we propose has two properties that are of high interest. Firstly, the 2DEG is contributed by ferro-electric doping, which is externally controllable, and thus tunable. Secondly, the hopping parameters between copper and oxygen in the system are found to be small, which is a desirable quality for cuprate superconductors.

The concept of this structure acts as a solid foundation to reach room temperature Tc, and one could easily envision similar systems with higher starting Tc and apply the same method to increase Tc as done here. Though room temperature is not observed in this study, we instead hope provide a novel and robust way to increase Tc from known systems.

In addition, the system exhibits two phases, which is in line with work being done with resistive switching being pursued with thin film oxides \cite{Choi}. Thin film oxides has long been a important area of discovery for many applications. The novelty of our case, where the interfacing of thin film oxide with ferro-electrics is key to tune the properties of the oxide, offers many new possibilities in addition to increasing Tc.

Currently, there have been successful methods of atomic layer deposition of layered lanthanum cuprates \cite{Sonsteby}. Key steps forward, such as fabrication of the full system via atomic layer deposition (ALD) or pulsed laser deposition (PLD), and whether further exploration into similar \XSL analogues will result in more controllable and higher Tc superconductors, remains to be seen.

\bibliographystyle{abbrv}
\bibliography{LaYBa}

\end{document}